\documentclass[journal]{IEEEtran}
\usepackage{amsmath,amsfonts,epsfig,graphicx}
\usepackage{subfig}
\usepackage{color,soul}
\usepackage{epstopdf}
\usepackage{tikz}
\usepackage{balance}
\usepackage{url}
\usepackage{multirow,hhline}
\usepackage{placeins}
\usepackage{wasysym}
\usepackage{amssymb}
\usepackage{pifont}
\usepackage[nomessages]{fp}
\usepackage{pgfplots}
\pgfplotsset{compat=newest}
\usetikzlibrary{plotmarks}
\usetikzlibrary{matrix}

\usepackage{tikz}

\newcommand{\FigPath}{./drawing/}
\newcommand{\image}[1]{\pgfuseimage{#1}}

\pgfdeclareimage[width = .95\columnwidth]{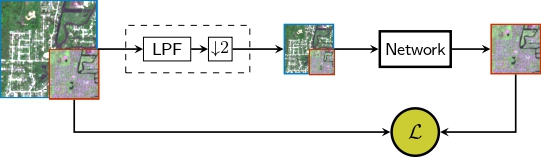}{\FigPath lowres.jpg}
\pgfdeclareimage[width = .95\columnwidth]{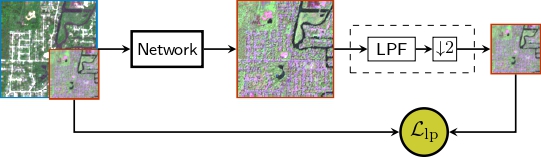}{\FigPath reversed.jpg}
\pgfdeclareimage[width = .95\columnwidth]{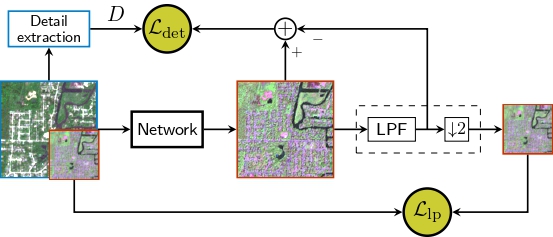}{\FigPath proposed.jpg}

\pgfdeclareimage[width = .36\columnwidth]{tempIm}{\FigPath matlab/X10.png}

\pgfdeclareimage[width = .36\columnwidth]{XtenA}{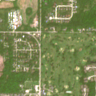}
\pgfdeclareimage[width = .36\columnwidth]{XtweA}{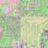}
\pgfdeclareimage[width = .36\columnwidth]{BaseA}{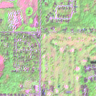}
\pgfdeclareimage[width = .36\columnwidth]{RevA}{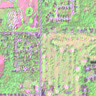}
\pgfdeclareimage[width = .36\columnwidth]{PropA}{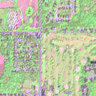}

\pgfdeclareimage[width = .36\columnwidth]{XtenB}{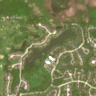}
\pgfdeclareimage[width = .36\columnwidth]{XtweB}{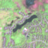}
\pgfdeclareimage[width = .36\columnwidth]{BaseB}{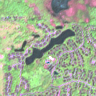}
\pgfdeclareimage[width = .36\columnwidth]{RevB}{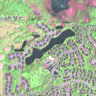}
\pgfdeclareimage[width = .36\columnwidth]{PropB}{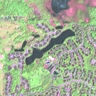}

\title{A full-resolution training framework for Sentinel-2 image fusion}

\author{Matteo Ciotola, Mario Ragosta, Giovanni Poggi, Giuseppe Scarpa
\thanks{
M.Ciotola and G.Poggi are with the Department of Electrical Engineering and Information Technology, University Federico II, Naples, Italy, e-mail: \{firstname.lastname\}@unina.it.
G.Scarpa is with the Engineering Department of the University Parthenope, Naples, Italy, e-mail: giuseppe.scarpa@uniparthenope.it.}
}
\begin{document}
%
\maketitle
\begin{abstract}
This work presents a new unsupervised framework 
for training deep learning models for super-resolution of Sentinel-2
images by fusion of its 10-m and 20-m bands.
The proposed scheme avoids the resolution downgrade process
needed to generate training data in the supervised case.
On the other hand,
a proper loss that accounts for cycle-consistency between the network prediction
and the input components to be fused is proposed.
Despite its unsupervised nature,
in our preliminary experiments
the proposed scheme has shown promising results 
in comparison to the supervised approach.
Besides,
by construction of the proposed loss,
the resulting trained network can be ascribed to the class of
multi-resolution analysis methods.
\end{abstract}
\section{Introduction}
\label{sec:intro}
Resolution enhancement of remotely sensed images 
is a low-level processing which is useful for many subsequent application-oriented 
tasks
such as target detection, change detection, forest and ice monitoring, 
land-cover classification and so forth.
Depending on the available imaging systems, 
it can assume different peculiar traits.
Except when a single resolution image is given,
in all other cases the super-resolution problem becomes a data fusion one.
A fusion that can be cross-sensor, multi-temporal, multi-resolution or a combination of the above.
In this work, 
we focus on the case of multi-resolution images acquired by a single imaging platform
which is Sentinel-2, 
with focus on the combination of the six 20-m spectral bands
with the four 10-m ones \cite{Gargiulo2019, Lanaras2018}.
The goal is therefore to raise the resolution of the former set of bands (20-m) to that of the latter set (10-m).
Contrarily to pansharpening \cite{Vivone2015, Vivone2020}, 
where a single high-resolution panchromatic band overlaps spectrally with the narrower-bands to be super-resolved,
here we dispose of four bands at target resolution of about 10-m which do not overlap spectrally with those to be super-resolved.
Nonetheless, 
it has been shown that the resolution enhancement benefits from the 10-/20-m data fusion,
although some 10-m bands weight more than others \cite{Gargiulo2018}.

In the last years,
deep learning solutions, notably convolutional neural networks (CNN), have assumed a key role 
in the context of resolution enhancement in remote sensing and beyond \cite{Lanaras2018, Masi2016, zhang2019pan, Cai2020},
due to the very promising results they have shown.
However, 
since original data do not come with ground-truth (GT),
researches who wanted to train any CNN solution 
had to find a proper way to run it. 
The most popular solution to this
is to resort to the so-called Wald's protocol,
also used for evaluation purposes,
which consists in a resolution downgrade of the available data
so that the original image bands to be super-resolved
could play as ground-truth instead (refer to \cite{Gargiulo2019} for further details),
whereas the scaled data become the input with which to feed the network 
during training. 
Examples of such an approach, first proposed by Masi {\em et al.} \cite{Masi2016}, 
are given in \cite{zhang2019pan, Cai2020, Hu2020, Deng2020, Peng2020, Zhang2020, Dong2020},
and specifically for the Sentinel-2 case in \cite{Gargiulo2019, Lanaras2018, Gargiulo2018, latte2020planetscope}.

Generally, these methods achieve very good scores in the same (reduced) resolution framework
where they have been trained and objective accuracy measurements can be obtained.
Instead, 
less clear-cut advantages are registered in the operational (full) resolution framework
where, unfortunately, only qualitative visual or numerical evaluations are possible.
An intuitive reason for which a performance shift presumably occurs moving to the operational frame
resides in the intrinsic informational gap between a given dataset and its corresponding 
``downgraded'', lower resolution, version used for training purposes.
In order to overcome this limitation several attempts have been made
resorting to different training paradigms, 
such as adversarial training or the use of a perceptual loss \cite{salgueiro2020super,pineda2020generative}.
Also, in \cite{Vitale2020} a cross-scale consistent training approach was proposed.
However, none of the above solutions makes a direct use of the full resolution images.
To the best of our knowledge, 
the only attempt in this regard is given in \cite{MA2020110}, 
where an adversarial training scheme is used. 
Rarely, 
it is possible to benefit from a complementary sensing system, 
flying closer to the ground,
which acquires nearly simultaneously the same scene of the target satellite using 
a (hopefully) identical imaging system.
In such a case, it could be possible to avoid the resolution downgrade process for training
as shown in \cite{galar2019super}.

\begin{figure*}
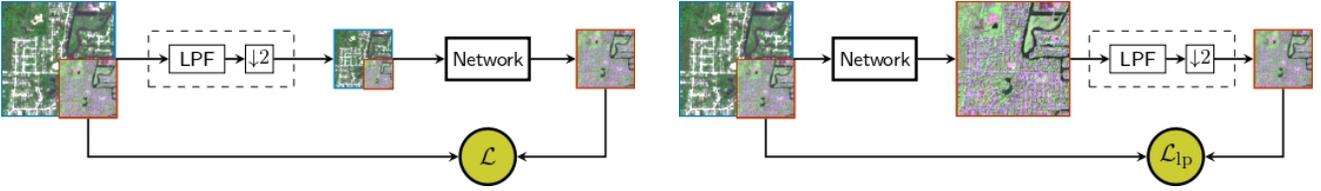

\setlength{\tabcolsep}{3mm}
\begin{tabular}{cc}
\image{lowres} &
\image{reversed} 
\end{tabular}
\caption{Conceptual schemes for training: baseline (left) and its
reversed (right).}
\label{fig:scheme}
\end{figure*}

\begin{figure}
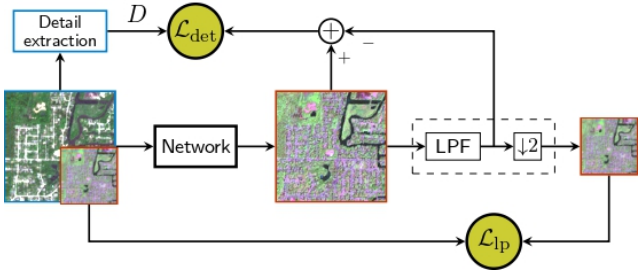

\centering
\image{proposed} 
\caption{Proposed training scheme with detail injection loss.}
\label{fig:proposed}
\end{figure}

In this work,
we propose a novel unsupervised paradigm for training at full resolution any Sentinel-2 super-resolution CNN.
To this aim we make use of the FUSE architecture proposed in \cite{Gargiulo2019}
as sample CNN model to highlight pros and cons of the proposed scheme.

The reminder of the paper is organized as follows. 
Next Section~\ref{sec:proposed} presents the proposed solution. 
Section~\ref{sec:exp} gathers and discusses our preliminary experimental results.
Finally, Section~\ref{sec:conclusion} draws conclusions 
discussing future perspectives.

\section{Proposed full-resolution training}
\label{sec:proposed}
The classic training paradigm for Sentinel-2 super-resolution networks
is summarized in the conceptual flowchart of Fig.~\ref{fig:scheme}~(left).
The available training sample image is properly scaled 
according with the estimated sensor's modulation transfer function (MTF).
This requires a low-pass filtering (LPF) followed by decimation.
The scaled sample is therefore ready to feed the network in order to adjust its parameters
according with some optimization schedule. 
The loss function to be optimized depends on the network prediction
and on the reference
given by the original (non-scaled) sample component to be super-resolved.

The basic motivation of the present proposal is that with this protocol the 
content of the original sample is not fully exploited because 
of the initial resolution downgrade process. 
Think of image features that emerge only in the 10-m resolution bands.
On the basis of the above consideration we decided to explore the possibility to 
perform the network prediction before scaling as depicted in Fig.~\ref{fig:scheme}~(right).
By doing so, 
the training is no longer supervised
since the network prediction is not compared to a reference truth but
further processed before to be compared to the input. 
Because of this
(numbers will be provided in the next section),
the simple inversion by itself does not provide improvements and can be detrimental for the accuracy.
This is easily justified by observing that the loss only sees 
a LPF version of the network prediction, 
therefore its high-pass content (detail) is out of control during training.

In order to put under control the detail component,
we added a dedicated loss term, 
with a reference properly extracted 
from the companion set of high-resolution bands.
This is summarized in the high-level flowchart of Fig.~\ref{fig:proposed},
where the overall loss is given by a linear combination of
the spectral ($\mathcal{L}_{\rm lp}$) and detail ($\mathcal{L}_{\rm det}$) loss components:
\[
\mathcal{L} = \mathcal{L}_{\rm lp} + \beta \mathcal{L}_{\rm det}
\]
The overall process underlies the hypothesis 
that all spectral bands have been normalized a priory.
As a consequence, 
at inference time, any image to be processed is first normalized
according to its own statistics which are then restored on the 
network output after prediction.
The way the loss is built allows us to also regard 
the proposed solution as belonging to the class 
of multiresolution analysis methods (MRA) \cite{Vivone2015}.
In fact, $\mathcal{L}_{\rm det}$ forces the network to inject a detail component into the prediction 
which is as close as possible to the provided reference detail $D$.
On the other hand, 
the spectral loss term 
ensures a cycle-consistency
which allows to keep under control eventual spectral distortions on the output.
Nonetheless, 
this new scheme remains unsupervised, likewise (b), 
and its comparison with the baseline solution
which is supervised has to be analysed with due care. 

The reference detail extraction proceeds as follows.
For each 20-m band $b$ we assess its local correlation coefficient 
with each of the 10-m bands (LPF versions), 
obtaining a correlation vector field $X^{(b)}=\{x_{i,j,k}^{(b)}\}$, 
being $(i,j)$ the spatial location and $k$ the selected 10-m band.
Then, a weight vector field $W^{(b)}=\{w_{i,j,k}^{(b)}\}$ is built using a softmax transform,
\[
w_{i,j,k}^{(b)} = \frac{\exp \left\{\gamma x_{i,j,k}^{(b)} \right\}}{\sum_h \exp\left\{\gamma x_{i,j,h}^{(b)}\right\}}, 
	\;\;\; \forall i, j,
\]
with a shrinking parameter $\gamma$.
Finally,
the pixel-wise weighted average of the 10-m details (high-pass components) yields 
the detail reference $D^{(b)}$ for $b$-th band.

\begin{table}
\centering
\footnotesize
\setlength{\tabcolsep}{4pt}
\begin{tabular}{lccccccc}
\hline
 & TA & Supervised &  $Q$ & $Q2^n$ & SAM & ERGAS & SCC \\
\hline
GT & & & 1 & 1 & 0 & 0 & 1\\
Pre-trained & n & y & 0.983 & 0.960 & 1.133 & 2.52 & 0.995  \\
Baseline & y & y & 0.983 & 0.969 & 1.067 & 2.12 & 0.993  \\
Reversed & y & n & 0.981 & 0.942 & 1.133 & 2.43 & 0.994 \\
Proposed & y & n & 0.982 & 0.946 & 1.116 & 2.37 & 0.995 \\
\hline
\end{tabular}
\caption{Numerical results. 
For each solution indicated in the leftmost column we report whether it is target-adaptive (TA) or not,
when it is supervised, and what is the accuracy according 
to several objective quality indicators \cite{Vivone2015}.}
\label{tab:results}
\end{table}

\begin{figure*}
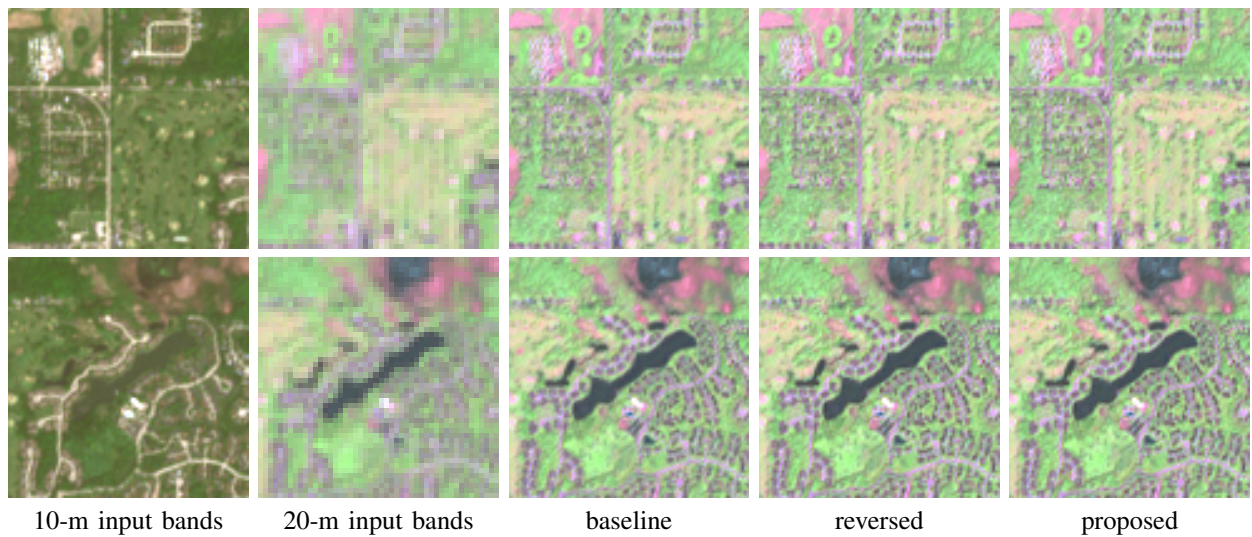

\centering
\setlength{\tabcolsep}{2pt}
\begin{tabular}{ccccc}
\image{XtenA} & \image{XtweA} & \image{BaseA} & \image{RevA} &\image{PropA} \\
\image{XtenB} & \image{XtweB} & \image{BaseB} & \image{RevB} &\image{PropB} \\
10-m input bands & 20-m input bands & baseline & reversed & proposed 
\end{tabular}
\caption{Visual comparison (sample crops) in false RGB representations. 
From left to right: 10-m and 20-m input components, outputs by baseline, reversed, and proposed.}
\label{fig:results}
\end{figure*}

\section{Experimental results}
\label{sec:exp}

In order to assess the behavior of the proposed training scheme
we assume the following experimental framework. 
A CNN architecture is selected (we use FUSE \cite{Gargiulo2019}).
Then, 
we (pre-)train it on a dataset reserved to this purpose,
obtaining a first reference method (pre-trained).
From this model we shift to the target-adaptive (TA) mode
as described in \cite{Scarpa2018a}.
Here,
a fixed number of training iterations are run on the same target image,
suitably arranged in a single mini-batch, to fine-tune the network on it.
On this training phase we apply the baseline approach of Fig.~\ref{fig:scheme}~(a)
and its reversed (b), and the proposed one of Fig.~\ref{fig:proposed}.

An objective numerical assessment is obtained by working in a reduced resolution framework
so that we can dispose of a GT for evaluation purposes.
In particular,
in this abstract paper we only show preliminary results (see Tab.~\ref{tab:results})
obtained on a single test image, 
deferring to the final full-paper and to the conference 
the presentation of additional ones.
The full-resolution version of the same test image 
is also used for a subjective visual comparison of the super-resolution results.
In Fig.~\ref{fig:results} are shown a couple of crops from the test image with 
related super-resolution results.

Numerical results highlight that the supervised approach (baseline) remains superior to the proposed unsupervised learning scheme in the reduced resolution framework. 
The ``reversed'' model looks less competitive than the proposed 
one, which includes a detail loss term and provides results closer
to those of the baseline. 
Moving to the full resolution frame, 
with the help of Fig.~\ref{fig:results}
we can appreciate (subjectively) by visual inspection
the detail enhancement level reached by the proposed approach
in comparison to the baseline.
Both the reversed model and the proposed, 
in fact, provide sharper structures and textures
without perceivable spectral distortions.

\section{Conclusion and perspectives}
\label{sec:conclusion}
In this work we have investigated a new training framework for 
Sentinel-2 super-resolution CNNs.
The proposed scheme allows to train the network on full resolution data, 
in an unsupervised manner,
hence avoiding the downgrade process needed to dispose of reference GT.
Results are encouraging but still below the level of a supervised scheme.
In particular,
the introduction of a deterministic detail generation process to complement the network loss function allows to almost fill the performance gap registered at reduced scale,
while achieving very good levels of sharpening at full scale.

The specific detail reference generation mechanism proposed here is just a first naïve solution
used as proof-of-concept which certainly deserves further study,
and we believe that there is a room left for improvement.
Finally, 
it is also worth underlying the unsupervised nature of the proposed scheme,
which makes it more robust with respect to the synthesis process 
to produce labeled data,
and the analogy with a well-established model-based technique for pansharpening,
that is the multiresolution analysis approach.



%

\balance
\bibliographystyle{IEEEtran}
\bibliography{refs}

\end{document}